# Analysis and Synthesis of the Disturbance Observer-based Robust Force Control Systems in State Space

Emre Sariyildiz
School of Mechanical, Materials, Mechatronic, and Biomedical Engineering,
Faculty of Engineering and Information Sciences
University of Wollongong
Wollongong, NSW, 2522, Australia
emre@uow.edu.au

*Abstract*— **This paper presents new analysis and synthesis tools for the robust force controllers implemented by Disturbance Observer (DOb) and Reaction Torque Observer (RTOb) in state space. The proposed design tool allows to use different plant and disturbance models in the observer synthesis, thus providing great flexibility in the robust force controller implementations. For example, the conventional DOb-based robust force controller can be synthesised using constant disturbance and interaction torque models in the DOb and RTOb synthesis. Moreover, the proposed tool allows to easily derive the transfer functions between the estimated interaction torque and exogenous reference, noise and disturbance inputs. This provides significant benefits when tuning the design parameters, such as the bandwidths of the observers, and analysing the stability and performance of the robust force controller. The proposed analysis method shows that the stability and performance of the robust force controller are significantly affected by the design parameters of the RTOb. For example, as the identified inertia is increased in the RTOb synthesis, an open-loop zero moves towards the right-half plane, leading to poor stability and performance limitations in practice. To synthesise a minimum-phase robust force controller, the identified inertia should be set smaller than or equal to the exact inertia. The stability and performance of the robust force control system are also affected by the other design parameters such as the bandwidths of the observers and environmental dynamics. To mitigate the latter effect, the design parameters of the force controller should be tuned by considering the dynamics of the environment that the servo system interacts physically. The proposed analysis and synthesis tools are verified by simulations.**

## I. INTRODUCTION

When robots physically interact with open and dynamic environments, the conventional robust position control systems fall-short in conducting high-performance motion control tasks in a safe manner [1–4]. In addition to the position state, the interaction forces (or torques) of servo systems should be accurately controlled to adjust the mechanical compliance for safe robot-environment interaction [5, 6]. Despite tremendous efforts, safe and high-performance physical robot-environment interaction still remains a great challenge for many next generation robotic systems, e.g., surgery robots, cobots, and exoskeletons [1, 4].

The safety of physical interaction tasks can be improved by using passive and active force control approaches [7–12]. The former approach can simply improve safety by reducing impact forces via soft and compliant mechanical components, yet they generally introduce several problems such as limited bandwidth, and complex dynamic models and position control problems [13–15]. In the latter approach, the compliance of a mechanical system is altered using direct aka "explicit" or indirect aka "implicit" active force controllers [16, 17]. In the explicit force control, the interaction force between the servo system and environment is controlled through the use of a force feedback controller (e.g., proportional, or proportional and integral controller) directly [16, 17]. On the other hand, indirect force controllers implicitly perform force control by deviating the motion of the servo system from the desired motion in accordance with the relation between the interaction force and adjustable control parameters of mechanical impedance and/or admittance [11, 16]. Although implicit force controllers can be implemented without measuring interaction forces in principal, an explicit force feedback loop is usually integrated into indirect force controllers to eliminate coupled impedance control and nonlinearity problems in practice [16]. Moreover, direct force control is essential for many physical robot environment interaction tasks, such as real-world haptics [18, 19]. Nevertheless, poor stability and performance are certain challenges for many direct force control systems, particularly when conducting tasks in open and dynamic environments [6].

To conduct high-performance interaction tasks, active force control systems have been widely studied in the last decades [7, 17]. Among them, the DOb based robust force controller proposed by Murakami and Ohnishi is one of the most popular force control techniques in the literature [20, 21]. The robust force controller is implemented using a cascade control structure and two observers, viz. DOb and RTOb. While the DOb improves not only the robustness but also the contact stability by suppressing disturbances in the inner-loop, the RTOb provides many benefits, e.g., force-sensorless force control and higher bandwidth of force estimation, in the outer-loop [22–25]. Moreover, the RTOb helps reduce the complexity of mechanical design and the cost of robotic systems [26]. However, the main drawback of the RTOb is that it is a model-based control method so the stability and performance of the force controller is affected by its design parameters [27–33]. Conventionally, it is assumed that the exact plant model is used in the design of the RTOb [19]. However, this assumption is impractical as it is not easy to obtain the exact dynamic model of many robotic systems. The conventional design approach also does not allow us to tune the stability and performance of the robust force controller using the RTOb's design parameters [6]. To synthesise high-performance robust force controllers, it is essential to understand how the stability and performance of the robust force controller change by the design parameters of the DOb (e.g., the bandwidth of the observer), RTOb (e.g., the identified inertia) and performance controller (e.g., force control gain).

To this end, this paper proposes new analysis and synthesis tools for the DOb-based robust force controllers in state space. To obtain the conventional DOb-based robust force controller, constant disturbance and interaction force models are respectively used in the DOb and RTOb syntheses in this paper. However, different disturbance models (e.g., periodic disturbances or dynamic disturbances for higher-order observer synthesis) can be similarly used in the proposed robust force controller synthesis. This paper also derives the transfer functions between the estimated interaction force and exogenous inputs using a compact equation. The transfer functions clearly show that the stability and performance of the robust force controller are significantly affected by the design parameters of the RTOb. For example, a non-minimum phase zero deteriorates the stability and performance of the robust force controller when the RTOb's identified inertia is not properly tuned. This paper also proposes practical tools that allow one to synthesise high-performance robust force controllers by properly tuning the force controller's design parameters, i.e., nominal inertia, identified inertia, bandwidths of the observers and force control gain. The proposed analysis and synthesis methods are verified by simulations.

The rest of the paper is organised as follows. Section II introduces DOb, RTOb and DOb-based robust force control system briefly. Section III proposes a new synthesis tool for the DOb-based robust force control systems in state space. Section IV analyses the stability and performance of the proposed robust force controller and presents simulation results. The paper ends with conclusion in Section V.

## II. DOb-based Robust Force Control Systems

The block diagrams of the DOb, RTOb and DOb-based robust force control system are illustrated in Fig. 1 [26]. In this figure, $J_m$, $J_{mn}$ and $J_{mi}$ represent the exact, nominal and identified inertiæ in the DOb and RTOb synthesis, respectively; $q_m$, $\dot{q}_m$ and $\ddot{q}_m$ represent the position, velocity and acceleration states of the motor, respectively; $\eta_{\dot{q}_m}$ represents the noise of the velocity measurement system; $D_{env}$ and $K_{env}$ represent the damping and stiffness of the environment that the servo system physically interacts, respectively; $\tau_d$ represents external disturbances, including the interaction torque $\tau_{int}$ and unknown disturbances $\tau_d^u$; $\tau_{int}^{ref}$ and $\hat{\tau}_{int}$ represent the reference and estimation of the interaction torque, respectively; $\tau_i$ represents the identified disturbances in the RTOb synthesis; $\hat{\tau}_{dis}$ represents the estimated internal and external disturbances by the DOb; $C_\tau$ represents the torque control gain; $g_{DOb}$ and $g_{RTOb}$ represent the bandwidths of the DOb and RTOb, respectively; and $\tau_m$ represents the control signal that is generated using the performance and robust control signals, i.e., $\tau_m^{per}$ and $\hat{\tau}_{dis}$, respectively.

Figure 1a shows that the DOb estimates internal and external disturbances using the nominal model of the servo system, control signal, velocity state and a low-pass-filter [26]. By

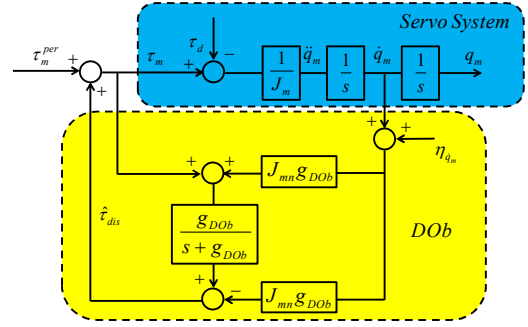

a) Block diagram of the DOb.

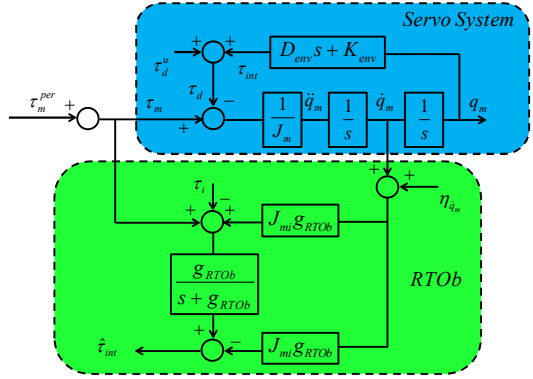

b) Block diagram of the RTOb.

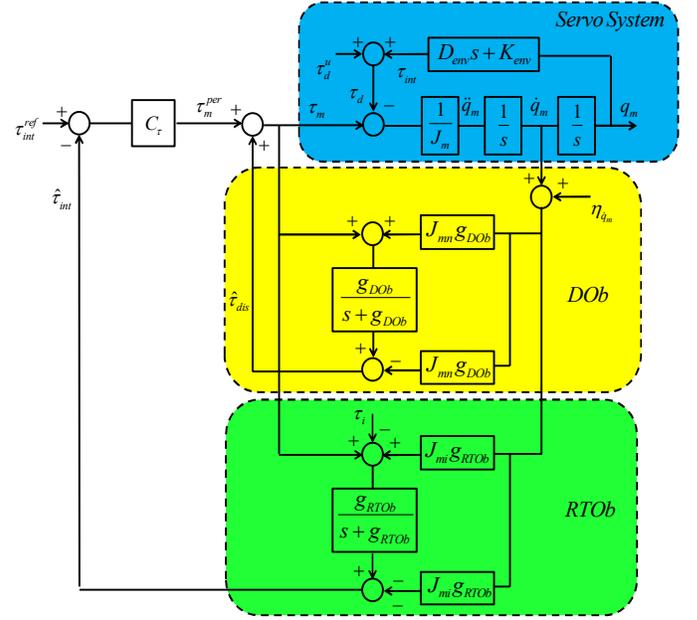

c) Block diagram of the DOb-based robust torque controller.

Fig. 1: Block diagrams of the DOb, RTOb and robust torque controller implemented using DOb.

identifying the model of the servo system accurately, the observer can be similarly used to estimate interaction forces as shown in Fig. 1b. This is generally called Reaction Torque Observer "RTOb" in the literature [20, 26].

As shown in Fig. 1c, the robust force controller is synthesised by employing two observers, DOb and RTOb, in the inner- and outer- loops, respectively. While the DOb improves the robustness of the force control system by feedbacking the estimated internal and external disturbances in the inner-loop, the force control performance is regulated through the use of a proportional force controller and the RTOb in the outer-loop. To estimate the interaction forces/torques, an RTOb is simply synthesised by adding the internal and external disturbances of a servo system to a DOb. Despite their similar control structures, the design parameters of the DOb and RTOb (e.g., the nominal and identified inertiæ $J_{mn}$ and $J_{mi}$, and the bandwidths of the observers $g_{DOb}$ and $g_{RTOb}$) have different effect on the stability and performance of the robust force controller. Moreover, the RTOb is a model-based controller, so the identified plant model may significantly affect the performance of force estimation. For example, while the stability of the robust force controller improves as the nominal inertia is increased in the DOb synthesis, higher values of the RTOb's identified inertia lead to more oscillatory force responses. To achieve high performance force control, not only the force control gain $C_\tau$, but also the design parameters of the DOb and RTOb ($J_{mn}, J_{mi}, g_{DOb}$ and $g_{RTOb}$) should be properly tuned.

For further details of the DOb-based robust force control systems, the reader is invited to refer to [6, 21].

III. DOB-BASED ROBUST FORCE CONTROLLER SYNTHESIS

To synthesise the DOb-based robust force controller, let us use the following general dynamic models in state space.

Exact Model: $\dot{x} = Ax + Bu - D\tau_d$ (1)

Nominal Model: $\dot{x} = A_n x + B_n u - D_n \tau_{dis}$ (2)

Identified Model: $\dot{x} = A_i x + B_i u + B_i \tau_i - D_i \tau_{int}$ (3)

Output: $y = x + \eta_x$ (4)

where $A, A_n$ and $A_i$ represent the state matrices; $B, B_n$ and $B_i$ represent the input vectors; $D, D_n$ and $D_i$ represent the disturbance vectors; $u = \tau_m$ represents the control input; $x = [q_m \ \dot{q}_m]^T$ represents the state vector; $y$ represents the output vector; and $\eta_x = [\eta_{q_m} \ \eta_{\dot{q}_m}]^T$ represents the noise vector of sensors at the output [34, 35].

A. DOb Synthesis in State Space

Let us assume that the dynamic model of the fictitious disturbance variable $\tau_{dis}$ is described using Eq. (5) and Eq. (6).

$$\dot{x}_D = A_D x_D \quad (5)$$

$$\tau_{dis} = C_D x_D \quad (6)$$

where $x_D$, $A_D$ and $C_D$ are the state vector, state matrix and output vector of the DOb, respectively.

Substituting Eq. (6) into Eq. (2) yields

$$\tilde{y}_D = A_n x + B_n u - \dot{x} = D_n C_D x_D \quad (7)$$

where $\tilde{y}_D$ is the output equation of the state vector $x_D$.

In practice, $\tilde{y}_D$ can be implemented using Eq. (8) due to the noise of measurement systems.

$$y_D = A_n (x + \eta_x) + B_n u - \dot{x} - \dot{\eta}_x \quad (8)$$

A closed-loop observer can be synthesised for the state vector $x_D$ using Eq. (5) and Eq. (8) as follows:

$$\begin{aligned}\dot{\hat{x}}_D &= A_D \hat{x}_D + L_D (y_D - D_n C_D \hat{x}_D) \\ &= (A_D - L_D D_n C_D) \hat{x}_{DOb} + L_D (A_n x + B_n u - \dot{x}) + L_D (A_n \eta_x - \dot{\eta}_x)\end{aligned} \quad (9)$$

where $L_D = [l_{D1} \ l_{D2}]^T$ is the DOb's gain vector that is yet to be tuned.

The observer given in Eq. (9) is impractical, because it comprises the derivative of the state vector $x$. This can be avoided by designing an observer for the auxiliary variable $\sigma = x_D + L_D x$.

An observer that estimates the auxiliary variable $\sigma$ can be synthesised by substituting $x_D = \sigma - L_D x$ into Eq. (9) as follows:

$$\dot{\hat{\sigma}} = A_\sigma \hat{\sigma} + A_{\sigma x} x + B_\sigma u + N_\sigma \eta \quad (10)$$

where $A_\sigma = A_D - L_D D_n C_D$, $A_{\sigma x} = L_D A_n + L_D D_n C_D L_D - A_D L_D$, $B_\sigma = L_D B_n$, $N_\sigma = [L_D A_n \ -L_D]$ and $\eta = [\eta_x \ \dot{\eta}_x]^T$.

The dynamics of the disturbance estimation error is derived by subtracting $\dot{\sigma}$ from Eq. (10) as follows:

$$\dot{e}_\sigma = A_\sigma e_\sigma + N_\sigma \eta \quad (11)$$

where $e_\sigma = \hat{\sigma} - \sigma$. The observer gain vector $L_D$ can be tuned using Eq. (11).

B. RTOb Synthesis in State Space

To synthesise the RTOb, let us assume that the dynamic model of the interaction torque $\tau_{int}$ is described using Eq. (12) and Eq. (13).

$$\dot{x}_R = A_R x_R \quad (12)$$

$$\tau_{int} = C_R x_R \quad (13)$$

where $x_R$, $A_R$ and $C_R$ are the state vector, state matrix and output vector of the RTOb, respectively.

Substituting Eq. (4) and Eq. (13) into Eq. (3) yields

$$y_R = A_i y + B_i u + B_i \tau_i - \dot{y} = D_i C_R x_R \quad (14)$$

where $y_R$ is the output equation of the state vector $x_R$.

An observer is synthesised for the state vector $x_R$ using Eq. (12) and Eq. (14) as follows:

$$\begin{aligned}\dot{\hat{x}}_R &= A_R \hat{x}_R + L_R (y_R - D_i C_R \hat{x}_R) \\ &= (A_R - L_R D_i C_R) \hat{x}_R + L_R (A_i x + B_i u + B_i \tau_i - \dot{x}) + L_R (A_i \eta_x - \dot{\eta}_x)\end{aligned} \quad (15)$$

where $L_R = \begin{bmatrix} l_{R1} & l_{R2} \end{bmatrix}^T$ is the RTOb's gain vector that is yet to be tuned.

To improve noise sensitivity, an auxiliary variable observer can be similarly synthesised by substituting $x_R = \rho - L_R x$ into Eq. (15) as follows:

$$\dot{\rho} = A_\rho \rho + A_{\rho x} x + B_\rho u + B_\rho \tau_i + N_\rho \eta \quad (16)$$

where $A_\rho = A_R - L_R D_I C_R$, $A_{\rho x} = L_R A_n - A_R L_R - L_R D_I C_R L_R$, $B_\rho = L_R B_n$ and $N_\rho = [L_R A_n \quad -L_R]$.

Similarly, the dynamics of the disturbance estimation error is derived by subtracting $\dot{\rho}$ from Eq. (16) as follows:

$$\dot{e}_\rho = A_\rho e_\rho + N_\rho \eta \quad (17)$$

where $e_\rho = \hat{\rho} - \rho$. Equation (17) can be used to tune the design parameters of the RTOb.

*C. DOb-based Robust Force Controller Synthesis*

The state space model of the robust force controller is obtained by combining Eq. (1), Eq. (10) and Eq. (16) as follows:

$$\dot{x}_a = A_{OL} x_a + B_a u + D_a \tau_d + B_{\rho a} \tau_i + N_a \eta \quad (18)$$

where $x_a = \begin{bmatrix} x \\ \hat{\sigma} \\ \hat{\rho} \end{bmatrix}$ and $A_{OL} = \begin{bmatrix} A & 0 & 0 \\ A_{\sigma x} & A_\sigma & 0 \\ A_{\rho x} & 0 & A_\rho \end{bmatrix}$ are the augmented state vector and its state matrix, respectively; $B_a = \begin{bmatrix} B & B_\sigma & B_\rho \end{bmatrix}^T$ and $B_{\rho a} = \begin{bmatrix} 0 & 0 & B_\rho \end{bmatrix}^T$ are the augmented input vectors; $D_a = \begin{bmatrix} D & 0 & 0 \end{bmatrix}^T$ and $N_a = \begin{bmatrix} 0 & N_\sigma & N_\rho \end{bmatrix}^T$ are the augmented disturbance and noise matrices, respectively.

Equation (18) represents the dynamic model of the open-loop control system. When the robustness of the force controller is improved by feedbacking the estimated disturbances in the inner-loop, i.e., $u = \tau_m^{per} + \hat{\tau}_{dis}$, Eq. (19) is obtained.

$$\dot{x}_a = A_{CLi} x_a + B_a \tau_m^{per} + D_a \tau_d + B_{\rho a} \tau_i + N_a \eta \quad (19)$$

where $A_{CLi} = \begin{bmatrix} A - BC_D L_D & BC_D & 0 \\ A_{\sigma x} - B_\sigma C_D L_D & A_\sigma + B_\sigma C_D & 0 \\ A_{\rho x} - B_\rho C_D L_D & B_\rho C_D & A_\rho \end{bmatrix}$ is the augmented state matrix. The other parameters are same as defined earlier.

When the robust force controller is implemented using a proportional performance controller, i.e., $\tau_m^{per} = r(t) - C_f \hat{\tau}_{int}$ where $r(t) = C_f \tau_{int}^{ref}$, as illustrated in Fig. 1, Eq. (20) is obtained.

$$\dot{x}_a = A_{CL} x_a + B_a r(t) + D_a \tau_d + B_{\rho a} \tau_i + N_a \eta \quad (20)$$

where $A_{CL} = \begin{bmatrix} A - BC_D L_D + C_f BC_R L_R & BC_D & -C_f BC_R \\ A_{\sigma x} - B_\sigma C_D L_D + C_f B_\sigma C_R L_R & A_\sigma + B_\sigma C_D & -C_f B_\sigma C_R \\ A_{\rho x} - B_\rho C_D L_D + C_f B_\rho C_R L_R & B_\rho C_D & A_\rho - C_f B_\rho C_R \end{bmatrix}$

is the augmented state matrix. The other parameters are same as defined earlier

IV. DOb-BASED ROBUST FORCE CONTROLLER ANALYSIS

Let us consider the following exact, nominal and identified servo system models in the analysis of the robust force controller.

$$\begin{aligned} J_m \ddot{q}_m &= \tau_m - \tau_d \\ J_{mn} \ddot{q}_m &= \tau_m - \tau_{dis} \\ J_{mi} \ddot{q}_m &= \tau_m - \tau_i - \left( \tau_{int} + \left( \tau_d^u - \tau_{id}^u \right) \right) \end{aligned} \quad (21)$$

where $\tau_d = \tau_d^u + D_{env} \dot{q}_m + K_{env} q_m$ is the external disturbance variable, including unknown disturbances and load torque due to physical interaction with environment; $\tau_{dis} = (J_m - J_{mn}) \ddot{q}_m + \tau_d$ is the fictitious disturbance variable including internal and external disturbances; and $\tau_i = (J_m - J_{mi}) \ddot{q}_m + \tau_{id}^u$ and $\tau_{id}^u$ are the identified internal and external disturbances used in the RTOb synthesis.

The state space models are obtained by substituting Eq. (21) into Eq. (1–4) as follows:

$$A_\bullet = \begin{bmatrix} 0 & 1 \\ 0 & 0 \end{bmatrix}, \quad B_\bullet = D_\bullet = \begin{bmatrix} 0 \\ 1/J_{m\bullet} \end{bmatrix} \quad (22)$$

where $\bullet$ is blank, $n$ and $i$ for the exact, nominal and identified servo system models, respectively.

Let us also assume that the disturbance and interaction torques are constant, i.e., $A_D = A_R = 0$. Substituting Eq. (22) into Eq. (11) and Eq. (17) yields

$$\dot{e}_\sigma = \left( -l_{D2}/J_{mn} \right) e_\sigma + N_\sigma \eta \quad (23)$$

$$\dot{e}_\rho = \left( -l_{R2}/J_{mi} \right) e_\rho + N_\rho \eta \quad (24)$$

The dynamics of estimation errors is adjusted by tuning the observer gains. For example, the conventional DOb and RTOb shown in Fig.1 are designed using $L_D = [0, g_{DOb} J_{mn}]$ and $L_R = [0, g_{RTOb} J_{mi}]$.

In the inner-loop, the closed-loop transfer functions of the robust force controller are derived using Eq. (25)

$$C_a \left( sI - A_{CLi} \right)^{-1} \left( B_a \tau_m^{per} + D_a \tau_d + B_{\rho a} \tau_i + N_a \eta \right) \quad (25)$$

where $C_a = \begin{bmatrix} 0 & -g_{RTOb} J_{mn} & 0 & 1 \end{bmatrix}$ is the output vector of the augmented system model. The transfer functions of the conventional DOb-based robust force controller are follows:

$$\hat{\tau}_{int} = L(s) \tau_m^{per} + LT_{\tau_d}(s) \tau_d + LT_{\tau_i}(s) \tau_i + LT_{\eta_{q_m}}(s) \eta_{q_m} \quad (26)$$

where $L(s) = \dfrac{g_{RTOb}}{s} \dfrac{s + g_{DOb}}{s + g_{RTOb}} \dfrac{(J_m - J_{mi}) s^2 + D_{env} s + K_{env}}{J_m s^2 + (g_{DOb} J_{mn} + D_{env}) s + K_{env}}$ represents the open-loop transfer function of the robust force controller;

$$T_{\tau_d}(s) = \frac{g_{RTOb}s^2(J_{mi}s + g_{DOb}J_{mn})}{g_{RTOb}(s+g_{DOb})((J_m - J_{mi})s^2 + D_{env}s + K_{env})}$$

represents the transfer function of the exogenous disturbance input;

$$T_{\tau_i}(s) = \frac{s(J_m s^2 + (g_{DOb}J_{mn} + D_{env})s + K_{env})}{(s+g_{DOb})((J_m - J_{mi})s^2 + D_{env}s + K_{env})}$$

represents the transfer function of the exogenous identified model input in the RTOb synthesis; and

$$T_{\eta_{\dot{q}_m}}(s) = \frac{g_{RTOb}s(J_{mi}s + g_{DOb}J_{mn})(J_m s^2 + D_{env}s + K_{env})}{g_{RTOb}(s+g_{DOb})((J_m - J_{mi})s^2 + D_{env}s + K_{env})}$$

represents the transfer function of the exogenous noise input.

Similarly, the closed-loop transfer functions of the robust force controller are derived by substituting Eq. (20) into Eq. (25) as follows:

$$\hat{\tau}_{int} = \frac{C_f L(s)}{1 + C_f L(s)} r + \frac{L T_{\tau_d}(s)\tau_d + L T_{\tau_i}(s)\tau_i + L T_{\eta_{\dot{q}_m}}(s)\eta_{\dot{q}_m}}{1 + C_f L(s)} \quad (27)$$

Equation (26) and Eq. (27) show that there is an integrator in the open-loop transfer function of the robust force controller. Therefore, the steady state error of force control goes to zero for a step input exponentially. There is also a phase-lead/lag compensator in the open-loop transfer function $L(s)$. This compensator can be tuned by adjusting the bandwidths of the DOb and RTOb. For example, a phase lead compensator is obtained by tuning $g_{RTOb} > g_{DOb}$, and the phase contribution of the compensator increases as the bandwidth of the RTOb (DOb) increases (decreases).

Equation (26) and Eq. (27) also show that the open-loop transfer function has a non-minimum phase zero when the RTOb is synthesised by tuning the identified inertia $J_{mi}$ higher than the exact inertia of the servo system $J_m$. This may lead to significant stability and performance problems by putting an upper bound on the force control gain $C_\tau$ in practice. Therefore, the RTOb should be synthesised using $J_{mi} \leq J_m$. As shown in these equations, the stability and performance of the robust force controller change not only by the control parameters $C_\tau, g_{DOb}$, $g_{RTOb}, J_{mn}$ and $J_{mi}$ but also by the environmental dynamics $D_{env}$ and $K_{env}$.

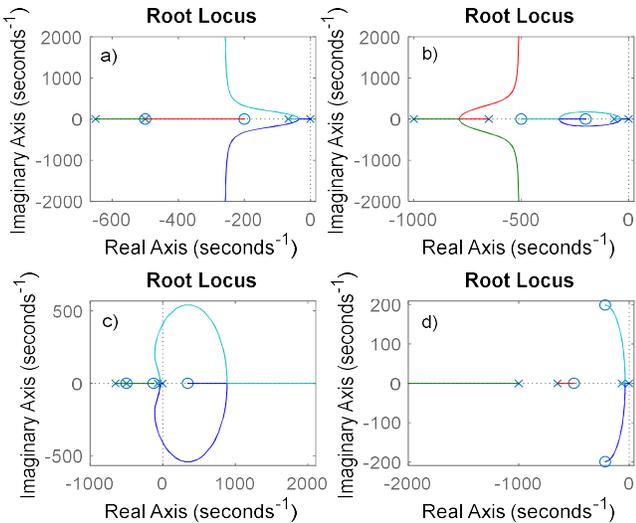

Fig. 2: Root-locus of the RTOb based robust force controller. $J_m = J_{mn} = 0.25$, $D_{env} = 50$ and $K_{env} = 10000$, a) $J_{mi} = J_m$ and $g_{DOb} = g_{RTOb} = 500$, b) $J_{mi} = J_m$ and $2g_{DOb} = g_{RTOb} = 1000$, c) $J_{mi} = 2J_m$ and $2g_{DOb} = g_{RTOb} = 1000$, and d) $J_{mi} = 0.5J_m$ and $2g_{DOb} = g_{RTOb} = 1000$.

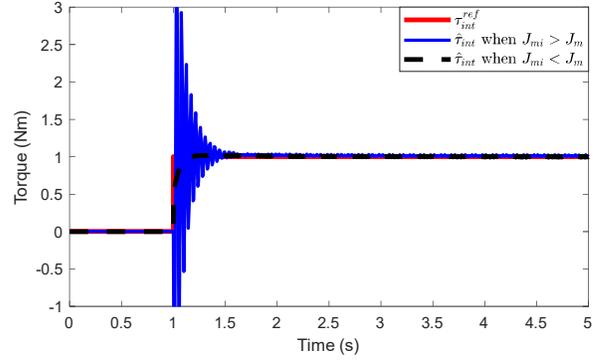

Fig. 3: Force control response when $J_m = J_{mn} = 0.25$, $D_{env} = 50$, $K_{env} = 10000$, $g_{DOb} = g_{RTOb} = 500$ and $C_f = 2$.

Figure 2 illustrates the root-loci of the robust force controller when different control parameters are used in the RTOb synthesis. Conventionally, the RTOb is designed using exact servo model, i.e., $J_{mi} = J_m$, and the same bandwidth values for the observers. The relative degree of the open-loop transfer function is two when $J_{mi} = J_m$, so a more oscillatory response is observed as the force control gain is increased (see Fig. 2a). The stability of the robust force controller can be improved by designing a phase lead compensator, i.e., by setting $g_{DOb} < g_{RTOb}$, as shown in Fig. 2b. When the identified inertia does not equal the exact inertia, the relative degree of the open-loop transfer function is one. However, the stability of the robust force controller notably changes by the identified inertia control parameter as shown in Fig. 2c and Fig. 2d.

Figure 3 illustrates the step responses of the robust force controller when the RTOb is tuned using different control parameters. As shown in this figure, the identified inertia of the RTOb should be set lower than the exact inertia to achieve good stability by avoiding the non-minimum phase zero in the open-loop transfer function $L(s)$.

## V. Conclusion

In this paper, new analysis and synthesis tools are presented for the DOb-based robust force control systems in state space. It is shown that the RTOb is a model-based control method, so the stability and performance of the robust force controller may significantly change when the design parameters of the RTOb are not properly tuned. For example, the non-minimum phase zero leads to poor stability and limited performance as shown in Fig. 2 and Fig. 3. In addition to the design parameters of the observers, the dynamics of environment affects the stability and performance of the robust force controller as shown in Section IV. To achieve good performance when the servo system interacts with open environments, adaptive controllers should be employed to tune the design parameters $C_\tau, g_{DOb}, g_{RTOb}, J_{mn}$ and $J_{mi}$ for different environmental dynamics.

The proposed design method allows to use different dynamic models of the servo system in the robust force controller synthesis. For example, the system matrices of Eq. (1–4) can be modified using Eq. (28) to integrate the viscous friction model of the servo system into the robust force controller.

$$A_\bullet = \begin{bmatrix} 0 & 1 \\ 0 & -b_{m\bullet}/J_{m\bullet} \end{bmatrix} \text{ and } B_\bullet = D_\bullet = \begin{bmatrix} 0 \\ 1/J_{m\bullet} \end{bmatrix} \quad (28)$$

where $b_{m\bullet}$ represents the exact, nominal and identified viscous friction coefficients when $\bullet$ is blank, $n$ and $i$, respectively. Similarly, we can use different models for the disturbance and interaction torques. This provides great flexibility to synthesise different DOb-based robust force controllers in practice.

Equation (10) and Eq. (16) are obtained by assuming that the velocity state of the servo system is measured in the implementation of the robust force controller. However, this is impractical for many robust force control applications, because the DOb and RTOb are often implemented by measuring the position state only [6, 27]. The dynamic of velocity estimation should be integrated into Eq. (10) and Eq. (16) to obtain a more realistic model for the robust force controller.

## REFERENCES

[1] A. Bicchi, M. A. Peshkin, J. E. Colgate, "Safety for physical human–robot interaction," B. Siciliano, and O. Khatib Ed., 1st ed. Springer Handbook of Robotics, Springer, Cham, 2016, pp 1335-1348.
[2] Barkan Ugurlu, et al., "Active Compliance Control Reduces Upper Body Effort in Exoskeleton-Supported Walking," *IEEE Trans. Human Machine Sys,* April 2020, vol. 50, no. 2, pp. 144-153.
[3] F. Bechet, et al., "Electro-Hydraulic Transmission System for Minimally Invasive Robotics", *IEEE Trans. on Ind. Electronics*, Dec. 2015, vol. 62, no. 12, pp. 7643-7654.
[4] C. C. Kemp, A. Edsinger and E. Torres-Jara, "Challenges for robot manipulation in human environments," *IEEE Robot. Autom. Mag.,*, vol. 14, no. 1, pp. 20-29, Apr. 2007.
[5] E. Sariyildiz, H. Temeltas, "Whole body motion control of humanoid robots using bilateral control," *Turk J Elec Eng & Comp Sci,* April 2017, vol. 25, no. 2, pp. 1495-1507.
[6] E. Sariyildiz, and K. Ohnishi, "An adaptive reaction force observer design," *IEEE/ASME Trans. Mechatron.*, vol. 20, no.2, pp. 750-760, Apr. 2015.
[7] G. Zeng, A. Hemami "An overview of robot force control," *Robotica*, vol. 15, no. 15, pp. 473–482, Sep. 1997.
[8] E. Sariyildiz, R. Mutlu, H. Yu "A Sliding Mode Force and Position Controller Synthesis for Series Elastic Actuators," Jan. 2020, vol. 38, no. 1, pp. 15-28.
[9] C. Tawk, E. Sariyildiz, G. Alici, "Force Control of a 3D Printed Soft Gripper with Built-in Pneumatic Touch Snsing Chambers pTSC," *Soft Robotics*, Oct. 2022, vol. 9, no. 5, pp. 970-980.
[10] F. Lange, W. Bertleff, M. Suppa, "Force and trajectory control of industrial robots in stiff contact," *In Proc. IEEE ICRA*, 2013 pp.2927-2934.
[11] N. Hogan, "Impedance Control: An Approach to Manipulation: Part I—Theory." ASME. *J. Dyn. Sys., Meas., Control*. March 1985; 107(1): 1–7.
[12] M. H. Raibert, and J. J. Craig, "Hybrid Position/Force Control of Manipulators." ASME. *J. Dyn. Sys., Meas., Control*. June 1981; 103(2): 126–133.
[13] M. Aydin, et al., "Variable Stiffness Improves Safety and Performance in Soft Robotics," *IEEE International Conference on Mechatronics (*ICM), Loughborough, United Kingdom, 15 – 17 March, 2023.
[14] E. Sariyildiz, C. Gong, H. Yu, "Robust Trajectory Tracking Control of Multi-mass Resonant Systems in State-Space," *IEEE Trans. Ind. Electron*, Dec. 2017, vol. 64, no. 12, pp. 9366 - 9377.
[15] B. Ugurlu, et al., "Benchmarking Torque Control Strategies for a Torsion based Series Elastic Actuator," *IEEE Robotics and Automation Magazine*, 2021.
[16] E. Sariyildiz, and K. Ohnishi, "On the explicit robust force control via disturbance observer," *IEEE Trans Ind. Electron.*, vol. 62, no. 3, pp. 1581-1589, Mar. 2015.
[17] L. Villani, J. D. Schutter, "Force Control," B. Siciliano, and O. Khatib Ed., 1st ed. Handbook of Robotics, Springer, Cham, 2016, pp 161-185.
[18] K. Ohnishi, S. Katsura and T. Shimono, "Motion Control for Real-World Haptics," in *IEEE Industrial Electronics Magazine*, vol. 4, no. 2, pp. 16-19, June 2010.
[19] K. Ohnishi, T. Mizoguchi, "Real haptics and its applications", *IEEJ Transactions on Electrical Engineering*, vol. 12, pp. 803-808, 2017.
[20] T. Murakami, F. Yu, and K. Ohnishi, "Torque sensorless control in multi-degree-of-freedom manipulator," IEEE Trans. Ind. Electron., vol. 40, no. 2, pp. 259–265, Apr. 1993.
[21] E. Sariyildiz, et al., "Discrete-Time Analysis and Synthesis of Disturbance Observer-based Robust Force Control Systems," *IEEE Access*, October 2021, vol. 9, pp. 148911 - 148924.
[22] S. Katsura, Y. Matsumoto, and K. Ohnishi, "Analysis and experimental validation of force bandwidth for force control," *IEEE Trans. Ind. Electron.*, vol. 53, no. 3, pp. 922–928, Jun. 2006.
[23] C. Mitsantisuk, et al., "Kalman filter-based disturbance observer and its applications to sensorless force control," *Adv. Robot.*, vol. 25, issue 3-4, pp. 335 – 353, Apr. 2011.
[24] S. Katsura, K. Irie, and K. Oishi, "Wideband force control by position-acceleration integrated disturbance observer," *IEEE Trans. Ind. Electron.*, vol. 55, no. 4, pp. 1699–1706, Apr. 2008.
[25] T. T. Phuong, K. Ohishi, and Y. Yokokura, "Fine sensorless force control realization based on dither periodic component elimination kalman filter and wide band disturbance observer," *IEEE Trans Ind. Electron*, vol. 67, no. 1, pp. 757-767, Jan. 2020.
[26] E. Sariyildiz and K. Ohnishi, "Stability and Robustness of Disturbance-Observer-Based Motion Control Systems," *IEEE Trans. Ind. Electron*, vol. 62, no. 1, pp. 414-422, Jan. 2015.
[27] E. Sariyildiz, et al., "Stability and Robustness of Disturbance Observer Based Motion Control Systems in Discrete Time Domain," *IEEE/ASME Trans. Mechatron*, August 2021, vol. 26, no. 4, pp. 2139-2150.
[28] H. Kobayashi, S. Katsura, and K. Ohnishi, "An analysis of parameter variations of disturbance observer for motion control," *IEEE Trans. Ind. Electron.*, vol. 54, no. 6, pp. 3413-3421, Dec. 2007.
[29] E. Sariyildiz, et al., "A stability analysis for the acceleration-based robust position control of robot manipulators via disturbance observer, *IEEE ASME Trans. Mechatron.*, vol. 23, no. 5, pp. 2369-2378, Oct. 2018.
[30] M. Bertoluzzo, G. S. Buja and E. Stampacchia, "Performance analysis of a high-bandwidth torque disturbance compensator," *IEEE/ASME Trans. on Mechatron.*, vol. 9, no. 4, pp. 653-660, Dec. 2004.
[31] K. Kong and M. Tomizuka, "Nominal model manipulation for enhancement of stability robustness for disturbance observer-based control systems," *International Journal of Control, Automation*, and Systems, vol. 11, no. 1, pp. 12-20, Feb. 2013.
[32] A. Tesfaye, H. S. Lee and M. Tomizuka, "A sensitivity optimization approach to design of a disturbance observer in digital motion control systems," *IEEE/ASME Trans. Mechatron.*, vol. 5, no. 1, pp. 32-38, 2000.
[33] E. Sariyildiz and K. Ohnishi, "Performance and robustness trade-off in disturbance observer design," *IECON 2013 - 39th Annual Conference of the IEEE Industrial Electronics Society*, 2013, pp. 3681-3686
[34] E. Sariyildiz, R. Mutlu, C. Zhang, "Active Disturbance Rejection Based Robust Trajectory Tracking Controller Design in State Space," *ASME Trans. J. Dyn. Sys., Meas., Control*, Jun 2019, vol. 141, no. 6, pp. 1-7 (061013).
[35] E. Sariyildiz "A Note on the Robustness of the Disturbance Observer-Based Robust Control Systems," *ASME Trans. J. Dyn. Sys., Meas., Control*, Sep 2022, vol. 144, no. 9, pp. 1-6 (094501).